\documentstyle[12pt]{article}

\begin{document}

\begin{center}
\null\vspace{2cm}
{\large {\bf Quantum Nonthermal Radiation of
Kerr-anti-de Sitter\\ Black Holes}}\\
\vspace{2cm}
M. Atiqur Rahman\footnote{$E-mail: atirubd@yahoo.com$}\\
{\it Department of Applied Mathematics,\\ University of Rajshahi,
Rajshahi - 6205, Bangladesh}
\end{center}
\vspace{3cm}
\centerline{\bf Abstract}
\baselineskip=18pt
\bigskip

We examine the properties of Quantum nonthermal radiation of a Kerr-anti-de Sitter (KAdS) black holes. Assuming that a crossing of the positive and negative Dirac energy levels occurs in a region near the event horizon of the hole, and spontaneous quantum nonthermal radiation
takes place in the overlap region. We solve the biquadratic equation governing the location of the event horizon of the KAdS black holes and present closed analytic expression for the radii of the horizons.\\
\vspace{0.5cm}
{\bf Keywords: Nonthermal radiation, Anti-de Sitter black holes.}\\
{\bf PACC: 9760L, 0420, 9720N, 6180J}
\vfill

\newpage

\section{Introduction}\label{sec1}

The quantum thermal radiation by black holes is an important discovery of Hawking \cite{one,two} that has been studied extensively in different types of spacetimes \cite{three,four,five,six,seven,eight,nine} including that of nonstatic \cite{ten,eleven,twelve} and nonstationary \cite{thirteen,fourteen} black holes. As black holes radiated quantum thermal radiation, there has a possibility to occurs also quantum nonthermal radiation by some of the black holes spacetime and recently, quantum nonthermal radiation has been investigated in the spacetimes of the nonstationary Kerr \cite{fifteen} and Kerr-Newman \cite{sixteen} black holes. So that, the study of the nonthermal radiation of black holes is interesting because it yields some new properties alongwith the results of some well-known black holes. In this paper we investigate the nonthermal radiation of a stationary KAdS black hole.

Anti-de Sitter (AdS) geometry has been considered as a challenging field for quantum field theory in different frameworks, including supersymmetry and string theory. The string /M-theory have also greatly stimulated the study of black hole solutions in AdS space. One example is the AdS/CFT correspondence between a weakly coupled gravity system in an AdS background and a strongly coupled conformal field theory (CFT) on its boundary \cite{seventeen,eighteen,nineteen}. The familiar Schwarzschild solution in AdS space describes a simple nonrotating AdS black hole that has a minimum critical temperature determined by the curvature radius of the AdS background. This implies that there must be a thermal phase transition between AdS space and SAdS space at a fixed temperature. This means that such black holes can be in stable equilibrium with thermal radiation at a certain temperature, while at temperatures higher than the critical value there is no stable equilibrium configuration without a black hole \cite{twenty}. It has been shown that small KAdS black holes become unstable, while large KAdS
black holes are always classically stable with respect to scalar and gravitational perturbations, via the superradiant amplification mechanism \cite{twenty one,twenty two,twenty three}.

Hawking et al. \cite{twenty four} have studied the relationship between KAdS black holes in the bulk and CFT living on a boundary Einstein universe. Probing the AdS/CFT correspondence the boundary Einstein universe that occurs at certain critical limit rotates at the speed of light. The authors also found that the generic thermodynamic features of the CFT agree with those of the black holes in the bulk. The simplest description of the boundary CFT in a rotating Einstein universe is imprinted by a KAdS black hole in the bulk and is very complicated and subtle question. The boundary Einstein universe that rotates at the speed of light due to the critical limit makes a significant simplification as it incorporates generic features of both bulk and boundary theories. So our study of KAdS black holes is necessary and meaningful.

The present paper is organization as follows. In section \ref{sec2}, we give a brief
description of the different spacetime of KAdS black holes. In section \ref{sec3}, we discuss the electromagnetic properties of KAdS black holes. In section \ref{sec4}, we investigate the quantum nonthermal effect of the KAdS black hole by using the Hamilton-Jacobi equation and generalized it using tortoise coordinates. Finally, in section \ref{sec5}, we present our conclusions.

\section{Spacetimes of KAdS Black Holes}\label{sec2}

The exact solution of the Einstein field equations with a cosmological constant that describes rotating black hole in asymptotically Anti-de Sitter behavior was given by Carter \cite{twenty five} of the form
\begin{eqnarray}
ds^2&=&\frac{1}{\Sigma }\left[\Delta _r -\Delta _\theta \,a^2\,{\rm sin^2\theta} \right]dt^2
+2\left[dt-a\,{\rm sin^2\theta} \,d\varphi \right]dr\nonumber\\
&&-\frac{\Sigma }{\Delta _\theta }d\theta ^2+\frac{2a}{\Sigma }\left[\Delta _\theta
(r^2+a^2)-\Delta _r \right]{\rm sin^2\theta} \,du\,d\varphi \nonumber\\
&&-\frac{1}{\Sigma }\left[\Delta _\theta (r^2+a^2)^2-\Delta _r \,a^2\,{\rm sin^2\theta}
\right]{\rm sin^2\theta} \,d\varphi^2\label{eq1},
\end{eqnarray}
where
\begin{eqnarray}
\Delta _r &=&(r^2+a^2)(1+\frac{r^2}{l^2})-2Mr,\nonumber\\
\Delta _\theta &=&1-\frac{a^2}{l^2}{\rm cos^2\theta} ,\hspace{1.3cm}
\Sigma =r^2+a^2\,{\rm cos^2\theta} ,\label{eq2}
\end{eqnarray}
$a$ is angular momentum and $M$ is the mass of the black hole, $l$ is the curvature radius related to the cosmological constant $\Lambda=-3l^{-2}$. The angular momentum parameter must satisfy the relation $a^2<l^2$, but when approach the critical value $a^2=l^2$, the metric becomes singular. In this limit the boundary of AdS spacetime is a three-dimensional Einstein universe which rotates at the speed of light.

For the inverse metric ${\rm g}^{\mu \nu }$ of Eq. (\ref{eq1}), we obtain
\begin{eqnarray}
\frac{\partial ^2}{\partial s^2}&=&-\frac{1}{\Sigma \Delta _\theta }
a^2\,{\rm sin^2\theta}\frac{\partial ^2}{\partial t^2}+\frac{2}{\Sigma}(r^2+a^2)\frac{\partial ^2}{\partial t\partial r}-\frac{2a}{
\Sigma \Delta _\theta }\frac{\partial ^2}{\partial t\partial \varphi }\nonumber\\
&&-\frac{\Delta _r }{\Sigma }\frac{\partial ^2}{\partial r^2}+\frac{2a}{\Sigma }\frac{\partial ^2}{\partial r\partial \varphi }-\frac{\Delta _\theta
}{\Sigma }\frac{\partial ^2}{\partial \theta ^2}-\frac{1}{
\Sigma \Delta _\theta \,{\rm sin^2\theta} }\frac{\partial ^2}{\partial \varphi ^2}\label{eq3},\\
{\rm g}&\equiv &{\rm det\rm g}_{\mu \nu }=-\Sigma ^2\,{\rm sin^2\theta} \label{eq4}.
\end{eqnarray}
We choose the component of the null tetrad vectors as \cite{twenty five}:
\begin{eqnarray}
l_\mu &=&\left\{1,\,0,\,0,\,-a\,{\rm sin^2\theta} \right\},\nonumber\\
n_\mu &=&\left\{\,\frac{\Delta _r}{2\Sigma },\,1,\,0,\,-\frac{1}{2\Sigma }\Delta _r \,\,a\,{\rm sin^2\theta} \right\},\nonumber\\
m_\mu &=&\frac{\sigma }{\sqrt{2}\,\rho }\left\{ia\,{\rm sin\theta} ,\,0,\,
-\frac{\Sigma }{\Delta _\theta },\,-i(r^2+a^2)
\,{\rm sin\theta} \right\},\nonumber\\
\bar m_\mu &=&\frac{\bar \sigma }{\sqrt{2}\,\bar \rho }\left\{
-ia\,{\rm sin\theta} ,\,0,\,-\frac{\Sigma }{\Delta _\theta },\,i(r^2+a^2)\,{\rm sin\theta} \right\}\label{eq5}.
\end{eqnarray}
The corresponding covariant quantities are
\begin{eqnarray}
l^\mu &=&\left\{0,\,1,\,0,\,0\right\},\nonumber\\
n^\mu &=&\left\{\frac{1}{\Sigma }(r^2+a^2),\,
-\frac{\Delta _r}{2\Sigma },\,0,\,\frac{a}{\Sigma
}\right\},\nonumber\\
m^\mu &=&\frac{1}{\sqrt{2}\,\rho \bar \sigma }\left\{ia\,{\rm sin\theta} ,\,0,\,\Delta _\theta ,\,\frac{i}{{\rm sin\theta} }\right\},\nonumber\\
\bar m^\mu &=&\frac{1}{\sqrt{2}\,\bar \rho \sigma }\left\{-ia\,{\rm sin\theta} ,\,0,\,\Delta _\theta ,\,-\frac{i}{{\rm sin\theta} }\right\}\label{eq6},
\end{eqnarray}
where
\begin{equation}
\rho =r+ia\,{\rm cos\theta} ,\hspace{1.3cm}\sigma
=1-\frac{i a}{l}\,{\rm cos\theta} \label{eq7},
\end{equation}
$\bar \rho $ and $\bar \sigma $ are complex conjugates of $\rho $ and $\sigma $,
respectively. The null tetrad (\ref{eq5}), (\ref{eq6}) satisfy the null, orthogonal
and metric conditions:
\begin{eqnarray}
l_\mu l^\mu &=&0,\hspace{1cm}n_\mu n^\mu =
0,\hspace{1cm}m_\mu m^\mu =0,\nonumber\\
l_\mu m^\mu &=&0,\hspace{1cm}n_\mu m^\mu =0,\hspace{1cm}l_\mu n^\mu
=1,\hspace{1cm}m_\mu \bar m^\mu =-1\label{eq8};\\
&&\nonumber\\
{\rm g}_{\mu \nu }&=&l_\mu n_\nu +l_\nu n_\mu
-m_\mu \bar m_\nu -m_\nu \bar m_\mu ,\nonumber\\
{\rm g}^{\mu \nu }&=&l^\mu n^\nu +l^\nu n^\mu
-m^\mu \bar m^\nu -m^\nu \bar m^\mu \label{eq9}.
\end{eqnarray}
The surface gravity is
\begin{equation}
\kappa =n_{\mu ;\nu }n^\mu l^\nu \label{eq10},
\end{equation}
which becomes with the help of Eqs. (\ref{eq5}) and (\ref{eq6})
\begin{equation}
\kappa =\frac{1}{2\Sigma }\left[\frac{\partial \Delta
_r }{\partial r}-\frac{2r}{\Sigma }\Delta _r\right]\label{eq11}.
\end{equation}
If $f=f(t,\,r,\,\theta ,\,\varphi )$ is a scalar function, then we have
\begin{equation}
\frac{df}{dn}=n^\mu \nabla _\mu f=n^\mu \partial
_\mu f,\hspace{1.3cm}\mu =0,\,1,\,2,\,3\label{eq12},
\end{equation}
which, for $f=f(t,\,r)$, becomes
\begin{equation}
\frac{df}{dn}=\left[n^0\partial _\mu +n^1\partial _r\right]f\label{eq13}.
\end{equation}
Obviously,
\begin{eqnarray}
\frac{dr}{dn}&=&-\frac{\Delta _r}{2\Sigma },\nonumber\\
&&\nonumber\\
\frac{d^2r}{dn^2}&=&-\left[n^0\partial _\mu +n^1\partial _r\right]
\frac{\Delta _r  }{2\Sigma }=-\frac{1}{2\Sigma ^2}(r^2+a^2)\frac{\partial \Delta _r
}{\partial t}+\frac{\Delta _r}{2\Sigma }\kappa \label{eq14}.
\end{eqnarray}
In the stationary case, the first term in the last equality of the second of eq. (\ref{eq14})
is equal to zero and determines the event horizons
\begin{equation}
\Delta _r =(r^2+a^2)(1+\frac{r^2}{l^2})-2Mr=0\label{eq15}.
\end{equation}

Next, we describe the horizon of the KAdS black holes. The radii of the horizons are determined by the real roots of the Eq. (\ref{eq15}). The Eq. (\ref{eq15}) can be rewritten as
\begin{equation}
r^4+(l^2+a^2)r^2-2l^2Mr+l^2a^2=0\label{eq16}.
\end{equation}
If $r_1,r_2,r_3,r_4$ are the roots of (\ref{eq16}), then we have
\begin{eqnarray}
r_1+r_2+r_3+r_4&=&0,\nonumber\\
r_1r_2+r_1r_3+r_1r_4+r_2r_3+r_2r_4+r_3r_4&=&a^2+l^2,\nonumber\\
r_1r_2r_3+r_1r_2r_4+r_2r_3r_4+r_1r_3r_4&=&2l^2M,\nonumber\\
r_1r_2r_3r_4&=&l^2a^2\label{eq17}.
\end{eqnarray}
The Eq. (\ref{eq17}) allow two real and a pair of complex conjugate roots. The largest of the real roots $r_1=r_+ $ corresponds to the radius of the hole\rq s outer event horizon and the other real root $r_2=r_- $ represents the radius of the inner Cauchy horizon. The real solution can be written from Eq. (\ref{eq16}) in a compact form using a real root $u$ of its resolvent equation of the form
\begin{eqnarray}
u=\frac{l^2+a^2}{3}+{\begin{array}{rl}&l^{4/3(M^2_{1e}-M^2_{2e})^{2/3}}\\
&(2N^2-M^2_{1e}-M^2_{2e})^{1/3}\end{array}}+l^{4/3}(2N^2-M^2_{1e}-M^2_{2e})^{1/3},\label{eq18}
\end{eqnarray}
where the two extreme mass parameters are given by
\begin{eqnarray}
M_{1e}=\frac{l}{\sqrt{54}}\sqrt{\zeta+\eta^3},\,\,M_{2e}=\frac{l}{\sqrt{54}}\sqrt{\zeta-\eta^3},\label{eq19}
\end{eqnarray}
with
\begin{eqnarray}
\zeta=\left(1+\frac{a^2}{l^2}\right)\left[\frac{36a^2}{l^2}
-\left(1+\frac{a^2}{l^2}\right)^2\right],\,\,\eta=\left[\left(1+\frac{a^2}{l^2}\right)^2+\frac{12a^2}{l^2}\right]^{1/2},\label{eq20}
\end{eqnarray}
and
\begin{equation}
N^2=M^2+\sqrt{(M^2-M^2_{1e})(M^2-M^2_{2e})}.\label{eq21}
\end{equation}
The mass parameter $M_{1e}$ has a definite physical meaning and for $l\rightarrow\infty$ we have $M^2_{1e}\rightarrow a^2$, but $M^2_{2e}\rightarrow -\infty$. The extreme mass $M_{1e}$ given in Eq. (\ref{eq19}) agrees with the value given in Ref. \cite{twenty}. It is clear that the black hole mass parameter $M$ must satisfy $M\geq M_{1e}$, that corresponds to an extreme black hole.
The horizons are located at the radii
\begin{eqnarray}
r_+=\frac{l}{2}(X+Y),\,\,r_+=\frac{l}{2}(X-Y),\label{eq22}
\end{eqnarray}
where
\begin{eqnarray}
X=\sqrt{u-l^2-a^2},\,\,Y=\sqrt{u-l^2-a^2+\frac{4Ml^2}{X}}.\label{eq23}
\end{eqnarray}
For $a=0$, the inner horizon vanishes and the outer horizon reduces to Schwarzschild-AdS black hole \cite{twenty six}. Expanding the expressions given in Eq. (\ref{eq22}) in powers of $1/l$ with $M/l\ll 1$, we get
\begin{eqnarray}
r_+&=&\tilde{r}_+ -\left(\frac{\tilde{r}^2_+}{2l^2}\right)\frac{2M\tilde{r}_+}{\tilde{r}_+-M}+O\left(\frac{1}{l^4}\right),\label{eq24}\\
r_-&=&\tilde{r}_- -\left(\frac{\tilde{r}^2_-}{2l^2}\right)\frac{2M\tilde{r}_-}{\tilde{r}_- -M}+O\left(\frac{1}{l^4}\right),\label{eq25}
\end{eqnarray}
where
\begin{equation}
\tilde{r}_\pm=M\pm\sqrt{M^2-a^2}.\label{eq26}
\end{equation}
From Eqs. (\ref{eq24}), (\ref{eq25}) and (\ref{eq26}) it is clear that the event horizon lies in the range $r_-<r_+<\tilde{r}_+$ and with $M=M_{1e}$, the outer and Cauchy horizons merge to form an extreme black hole with radius
\begin{equation}
r_{ch}=\frac{l}{\sqrt{6}}\left(\eta-1-\frac{a^2}{l^2}\right)^{1/2}.\label{eq27}
\end{equation}
In the critical limit of rotation i.e., $a^2=l^2$, from Eqs. (\ref{eq22}) and (\ref{eq27}) we get the limiting size for the extreme black hole horizon as
\begin{equation}
\tilde{r}_{ch}=\frac{3}{\sqrt{8}}M=\frac{l}{\sqrt{3}},\label{eq28}
\end{equation}
but the angular velocity becomes infinity i.e., the boundary of Einstein universe rotating at the speed of light.

\section{Electromagnetic Properties of KAdS black holes}\label{sec3}
Here we now consider that the KAdS black holes described above may also posses a small test electric charge. The obvious thing to do is to seek a electromagnetic potential $A$ such that the Hamilton-Jacobi theory or a physical point of view by the correspondence principle ensuring not only geodesics but also closed particles orbits will be integrable. Interms of an electromagnetic field potential
\begin{equation}
A=A_\mu dx^\mu,\label{eq29}
\end{equation}
the motion of a particle with charge $e$ and mass $m$ in stationary KAdS spacetime with metric tensor
${\rm g}^{\mu \nu }$ is given by the Hamilton-Jacobi equation \cite{twenty seven}
\begin{equation}
{\rm g}^{\mu \nu }\left(\frac{\partial S}{\partial x^\mu }-eA_\mu \right)
\left(\frac{\partial S}{\partial x^\nu }-eA_\nu \right)-m^2=0,\label{eq30}
\end{equation}
where $S(t,\,r,\,\theta ,\,\varphi )$ is the Hamilton principal function, and $A_\mu $
is the four-potential.
In order to introduce the necessary cross terms for KAdS black hole case, we shall start off by requiring that the electromagnetic field potential $A$ contain only the same two components as were necessary in the spherical case of the form
\begin{equation}
A=A_0 dt+A_3 d\varphi,\label{eq31}
\end{equation}
where $A_0$ and $A_3$ are scalar, independent of $t$ and $\varphi$.
The effect of the electromagnetic field of this charge on this spacetime geometry of KAdS black holes can be neglected and the spacetime can still be well described by the metric given in Eq. (\ref{eq1}). The Associated solution of the source free Maxwell\rq s equations in asymptotically flat case is constructed using the well known fact that for Ricci-flat metrics a Killing one-form is closed and co-closed \cite{twenty eight,twenty nine}. Although the KAdS metric, we have consider here is not Ricci-flat, one can still use the killing isometries to describe the electromagnetic field of a KAdS black hole \cite{thirty}. The desired potential one-form is given by
\begin{equation}
A=-\frac{Qr}{\Sigma}\left(dt-\frac{a\,{\rm sin^2 \theta}}{1-a^2/l^2}d\varphi\right),\label{eq32}
\end{equation}
where the parameter $Q$ is related to the electric charge $Q'=Q/(1-a^2/l^2)$ of the black hole given by the Gaussian flux
\begin{equation}
Q'=\frac{1}{4 \pi}\oint ^* F,\label{eq33}
\end{equation}
where
\begin{equation}
F=\frac{Q(\Sigma-2r^2)}{\Sigma^2}\left(dt-\frac{a\,{\rm sin^2 \theta}}{1-a^2/l^2}d\varphi\right)\wedge dr+\frac{Qra{\rm sin2\theta}}{\Sigma^2}\left(adt-\frac{r^2+a^2}{1-a^2/l^2}d\varphi\right)\wedge d\theta.\label{eq34}
\end{equation}

\section{Quantum Nonthermal Effect}\label{sec4}
The Hamilton-Jacobi equation (\ref{eq30}) with Eqs. (\ref{eq3}) and  (\ref{eq31}) gives
\begin{eqnarray}
&&\nonumber\\
\frac{a^2\,{\rm sin^2\theta} }{\Delta _\theta }
\left(\frac{\partial S}{\partial t}\right)^2
&+&\Delta _r \left(\frac{\partial S}{\partial r}\right)^2
+\Delta _\theta \left(\frac{\partial S}{\partial \theta }\right)^2
+\frac{1}{\Delta _\theta \,{\rm sin^2\theta} }
\left(\frac{\partial S}{\partial \varphi }\right)^2\nonumber\\
&&\nonumber\\
&-&2(r^2+a^2)\frac{\partial S}{\partial t}
\frac{\partial S}{\partial r}+\frac{2a}{\Delta _\theta }
\frac{\partial S}{\partial t}\frac{\partial S}{\partial \varphi }
-2a\frac{\partial S}{\partial r}
\frac{\partial S}{\partial \varphi }\nonumber\\
&&\nonumber\\
&-&\frac{2e}{\Delta _\theta }(a^2A_0\,{\rm sin^2\theta} +aA_3)
\frac{\partial S}{\partial t}+2e
\left[(r^2+a^2)A_0+aA_3\right]\frac{\partial S}{\partial r}\nonumber\\
&&\nonumber\\
&-&\frac{2e}{\Delta _\theta }\left(aA_0+\frac{A_3}{{\rm sin^2\theta} }\right)
\frac{\partial S}{\partial \varphi }
+\frac{a^2\,{\rm sin^2\theta} }{\Delta _\theta }e^2{A_0}^2
+\frac{1}{\Delta _\theta \,{\rm sin^2\theta} }e^2{A_3}^2\nonumber\\
&&\nonumber\\
&+&\frac{2ae^2}{\Delta _\theta }A_0A_3
+m^2(r^2+a^2\,{\rm cos^2\theta} )=0\label{eq35}.
\end{eqnarray}
Introducing the generalized tortoise coordinates \cite{thirty one} of the following form
\begin{eqnarray}
r_*&=&r+\frac{1}{2\kappa}{\rm ln}(r-r_+),\nonumber\\
t_*&=&t,\nonumber\\
\theta_*&=&\theta,\label{eq36}
\end{eqnarray}
we have
\begin{eqnarray}
\frac{\partial }{\partial r}&=&\left(1+\frac{1}{2\kappa(r-r_+)}\right)\frac{\partial }{\partial r_*},\nonumber\\
\frac{\partial }{\partial t}&=&\frac{\partial }{\partial t_*},\nonumber\\
\frac{\partial }{\partial \theta}&=&\frac{\partial }{\partial \theta_*},\label{eq37}
\end{eqnarray}
where $r_+$ is the event horizon of KAdS black hole.
The spacetime admits a Killing vector $(\partial S/\partial \varphi )^\alpha $, so that
\begin{equation}
\frac{\partial S}{\partial \varphi }=\chi \;\;{\rm (constant)}.\label{eq38}
\end{equation}
Let us define
\begin{equation}
S=T(t_*,r_*,\theta_*)+\chi \varphi,\label{eq39}
\end{equation}
and
\begin{equation}
\omega =-\frac{\partial S}{\partial t_*},\hspace{1.0cm}
\ell=\frac{\partial S}{\partial \theta_*},\label{eq40}
\end{equation}
where $\omega$ is the energy and $\ell$ is the angular momentum of a particle.
Using Eqs. (\ref{eq36})-(\ref{eq40}), we can write  Eq. (\ref{eq35}) of the form
\begin{equation}
P\left(\frac{\partial S}{\partial r_*}\right)^2
-4\kappa(r-r_+)R\left(\frac{\partial S}{\partial r_*}\right)
+\left[2\kappa(r-r_+)\right]^2T=0,\label{eq41}
\end{equation}
where
\begin{eqnarray}
P&=&\Delta _r (2\kappa(r-r_+)+1)^2,\nonumber\\
R&=&[2\kappa(r-r_+)+1]a\chi-(r^2+a^2)
[2\kappa(r-r_+ +1]\omega-e\left[(r^2+a^2)A_0+aA_3][2\kappa(r-r_+)+1\right],\nonumber\\
T&=&\frac{a^2\,{\rm sin^2\theta} }{\Delta _\theta }\omega ^2
+\Delta _\theta \,\ell^2+\frac{1}{\Delta _\theta \,{\rm sin^2\theta} }\chi ^2
-\frac{2a}{\Delta _\theta }\omega \chi +\frac{2e}{\Delta _\theta }\left(a^2A_0\,{\rm sin^2\theta} +aA_3\right)\omega \nonumber\\
&&-\frac{2e}{\Delta _\theta }\left(aA_0+\frac{A_3}{{\rm sin^2\theta} }\right)\chi
+\frac{1}{\Delta _\theta }\left(a^2\,{\rm sin^2\theta} \,e^2{A_0}^2
+\frac{1}{{\rm sin^2\theta} }e^2{A_3}^2+2ae^2A_0A_3\right)\nonumber\\
&&+m^2\left(r^2+a^2\,{\rm cos\theta} \right).\label{eq42}
\end{eqnarray}
The solutions of Eq. (\ref{eq41}) are given by
\begin{equation}
\frac{\partial S}{\partial r_*}=\frac{2\kappa(r-r_+)}{P}\left(R\pm
\sqrt{R^2-PT}\right)\label{eq43}.
\end{equation}
Since $S$ and $\partial S/\partial r_*$ cannot be imaginary number, we have
\begin{equation}
R^2-PT\geq 0\label{eq44}.
\end{equation}
From the equality given in Eq. (\ref{eq44}) we have
\begin{equation}
\omega ^\pm =\frac{-W_2\pm \sqrt{{W_2}^2-W_1W_3}}{W_1},\label{eq45}
\end{equation}
where
\begin{eqnarray}
&&W_1=\left((r^2+a^2)[2\kappa(r-r_+)+1]\right)^2-\frac{a^2\,{\rm sin^2\theta} }{\Delta _\theta }P,\nonumber\\
&&W_2=\left((r^2+a^2)[2\kappa(r-r_+)+1]\right)\Big(e[(r^2+a^2)A_0+aA_3][2k(r-r_+)+1]\nonumber\\
&&-a\chi [2\kappa(r-r_+)+1]\Big)+\frac{1}{\Delta _\theta }P[a\chi -e(a^2A_0\,{\rm sin^2\theta} +aA_3)],\nonumber\\
&&W_3=\left(e[(r^2+a^2)A_0+aA_3][2\kappa(r-r_+)+1]-a\chi [2\kappa(r-r_+)+1]\right)^2\nonumber\\
&&-P\left[\Delta _\theta \,\ell^2+\frac{\chi^2}
{\Delta _\theta \,{\rm sin^2\theta} }-\frac{2e}{\Delta _\theta }
\left(aA_0+\frac{A_3}{{\rm sin^2\theta} }\right)\chi \right.\nonumber\\
&&+\frac{1}{\Delta _\theta }\left(a^2\,{\rm sin^2\theta} \,e^2{A_0}^2
+\frac{1}{{\rm sin^2\theta} }e^2{A_3}^2+2ae^2A_0A_3\right)+m^2\left.\left(r^2+a^2\,{\rm cos^2\theta} \right)\right].\label{eq46}
\end{eqnarray}
Considering the sign of inequality given in Eq. (\ref{eq44}), the distribution of the energy levels of the Dirac vacuum is given by
\begin{equation}
\omega \geq \omega ^+\qquad \mbox{and}\qquad
\omega \leq \omega ^-.\label{eq47}
\end{equation}
For stationary KAdS spacetime, the distribution of the energy levels of the Dirac vacuum is described exactly by Eqs. (\ref{eq45})-(\ref{eq47}), but the forbidden region is
\begin{equation}
\omega ^-<\omega <\omega ^+\label{eq48}.
\end{equation}
The width of the forbidden region is
\begin{equation}
\Delta \omega =\omega ^+-\omega ^-
=2\frac{\sqrt{{W_2}^2-W_1W_3}}{W_1}\label{eq49}.
\end{equation}
For $r\rightarrow \infty $, we have
\begin{equation}
\omega ^\pm \rightarrow \pm m\label{eq50}.
\end{equation}
In this case, the distribution of the Dirac energy levels reduces to that in the
Minkowski spacetime. The width of the forbidden region becomes $\Delta \omega =2m$.

We now consider the case near the event horizon, i.e., when $r\rightarrow r_+$.
We have from Eq. (\ref{eq41})
\begin{eqnarray}
\lim_{r\rightarrow r_+}P={\Delta _{r_+}}=0\label{eq51},
\end{eqnarray}
where
$$
\Delta _{r+}=(r_+^2+a^2)(1+\frac{r_+^2}{l^2})-2Mr_+.
$$
This is just the null surface condition, so the limit of $P$ is zero. With Eq.
(\ref{eq51}) we get from Eqs. (\ref{eq45}) and (\ref{eq46}) that
\begin{eqnarray}
&&\lim_{r\rightarrow r_+}\left({W_2}^2-W_1W_3\right)=0,\label{eq52}\\
&&\nonumber\\
\omega _0&=&\lim_{r\rightarrow r_+}\omega ^+=
\lim_{r\rightarrow r_+}\omega ^-=-\lim_{r\rightarrow r_+}\frac{W_2}{W_1}\nonumber\\
&=&\frac{a\,\chi-e\Phi}{(r_+^2+a^2)},\label{eq53}
\end{eqnarray}
with
$$
\Phi =\left[(r_+^2+a^2)A_0+aA_3\right].
$$
Then Eq. (\ref{eq49}) indicates that the width of the forbidden region vanishes at the
event horizon, and there exists a crossing of the positive and negative
energy levels near the event horizon. For $\omega _0>+m$, the particle can escape to
infinity from the event horizon. That is, there occurs the Starobinsky-Unruh process
(spontaneous radiation) in the region near the event horizon. For the stationary KAdS spacetime, it is very interesting that the maximum energy $\omega_0$ of a particle depends on the horizon shape and on the four-potential $A$ i.e., the test electric charge of the KAdS black hole. The maximum energy of a particle in this quantum nonthermal effect is $\omega_0$ \cite{thirty two,thirty three,thirty four,thirty five}. While the energy extended by the radiation particle is \cite{thirty six,thirty seven,thirty eight}
\begin{eqnarray}
m<\omega\leq\omega _0=\frac{a\,\chi-e\Phi}{(r_+^2+a^2)}.\label{eq54}
\end{eqnarray}
This radiation is independent of temperature and is a nonthermal radiation that differ from the Hawking\rq s radiation \cite{thirty nine,fourty,fourty one,fourty two,fourty three}.
For KAdS black hole, the dragging velocity of the outer horizon is given by
\begin{equation}
\Omega_+=\frac{a(1-\frac{a^2}{l^2})}{(r^2_+ +a^2)}\label{eq55}.
\end{equation}
Using Eq. (\ref{eq50}) in Eq. (\ref{eq48}), we get
\begin{equation}
\omega_0=\frac{\Omega_+ (\chi-e\Phi/a)}{(l^2-a^2)}\label{eq56}.
\end{equation}
In the critical limit of rotation (i.e., $a^2=l^2$), the outer and inner horizon merge to form an extreme black hole with radius given in Eq. (\ref{eq28}). In this case, the maximum energy of the Dirac particles that crossing the boundary of the Einstein universe rotating at the speed of light is infinite. That is, escape the boundary with the speed of light.

\section{Conclusions}\label{sec5}

We have studied nonthermal radiation of a stationary KAdS black hole assuming that the black holes may carry a test electric charge such that the Killing one-form can be used as a potential one-form \cite{thirty} for the associated electromagnetic field to introduce necessary cross terms for KAdS black hole case. We have also solved the biquadratic equation to determine the position of different KAdS black hole event horizons and found the radii of the horizons in analytical formulas.

Here, we have derived the exact expressions of the energy for the positive and negative states. The positive energy state is interlaced by the negative energy state in a region near the event horizon of the KAdS black hole so that there occurs a spontaneous nonthermal radiation (Starobinsky-Unruh process) in the overlap region. The formulae and results are formally the same as for the nonstationary black hole. For nonstationary KAdS black hole, the only difference is that $M$ is the function of retarded time $t$ while in the stationary case $M$ is constants.

The important result, we have found here is that when the Einstein universe rotates at the speed of light, the Dirac particles crossing the boundary of Einstein universe escape with the speed of light. The results of our work correspond to the nonstationary Kerr \cite{fifteen} black hole when $\Lambda =0$ and the black hole mass is the function of retarded time $t$ and  nonstationary Kerr-Newman \cite{sixteen} black hole when $\Lambda =0$; $M$ and $Q$ are the functions of retarded time $t$. So our work on the stationary KAdS black hole is well motivated.

\vspace{0.5cm}
\noindent
{\large\bf Acknowledgement}\\
The author (MAR) thanks the Abdus Salam International Centre for Theoretical Physics (ICTP), Trieste, Italy, for giving opportunity to utilize its e-journals for research purpose.


\begin{thebibliography}{99}
\bibitem{one}
Hawking, S.W. (1974). {\em Nature} {\bf 248}, 30
\bibitem{two}
Hawking, S.W.  (1975). {\em Commun. Math. Phys.} {\bf 43}, 199.
\bibitem{three}
Gibbons, G.W. and Hawking, S.W. (1977). {\em Phys. Rev. D} {\bf 15}, 2738.
\bibitem{four}
Liu Liao and Dian-Yan Xu, (1980). {\em Acta Phys. Sin.} {\bf 29}, 1617.
\bibitem{five}
Zhao Zheng, Yuan-Xing Guei and Liu Liao, (1981).
{\em Acta Astrophys. Sin.} {\bf 1}, 141.
\bibitem{six}
Dai Xianxin, Zhao Zheng and Liu Liao, (1993). {\em Science in China A}
{\bf 23}, 69.
\bibitem{seven}
Dai Xianxin and Zhao Zheng, (1992). {\em Acta Phys. Sin.} {\bf 41}, 868.
\bibitem{eight}
Ahmed, M. (1987). {\em Class. Quantum Grav.} {\bf 4}, 431.
\bibitem{nine}
Ahmed, M. and Mondal, A.K. (1993). {\em Phys. Lett. A} {\bf 184}, 37.
\bibitem{ten}
Li Zhongheng and Zhao Zheng, (1993). {\em Chin. Phys. Lett.} {\bf 10}, 126.
\bibitem{eleven}
Zhao Ren, Han Fulong and Wu Yueqin, (1997). {\em Il Nuovo Cimento B}
{\bf 112}, 701.
\bibitem{twelve}
Zhang Lichun, Wu Yueqin and Zhao Ren, (1999). {\em Il Nuovo Cimento B}
{\bf 114}, 1163.
\bibitem{thirteen}
Li Zhongheng and Zhao Zheng, (1995). {\em Science in China (Series A)}
{\bf 38}, 74.
\bibitem{fourteen}
Jing Jiliang and Wang Yongjiu, (1997). {\em Intl. J. Theor. Phys.}
{\bf 36}, 1745.
\bibitem{fifteen}
Yang Shuzheng and Zhao Zheng, (1996). {\em Intl. J. Theor. Phys.}
{\bf 35}, 2455.
\bibitem{sixteen}
Junli Lu, (1999). {\em Intl. J. Theor. Phys.} {\bf 38}, 2029
\bibitem{seventeen}
J. M. Maldacena, (1998). {\em Adv. Theor. Math. Phys.} {\bf 2}, 231.
\bibitem{eighteen}
S. S. Gubser, I. R. Klebanov and A. M. Polyakov, (1998). {\em Phys. Lett. B} {\bf 428}, 105.
\bibitem{nineteen}
E. Witten, (1998). {\em Adv. Theor. Math. Phys.} {\bf 2}, 253.
\bibitem{twenty}
S. W. Hawking and D. Page, (1983). {\em Commun. Math. Phys.} {\bf 87}, 577.
\bibitem{twenty one}
S. W. Hawking and H. S. Reall, (1999). {\em Phys. Rev. D} {\bf 61}, 024014.
\bibitem{twenty two}
V. Cardoso and O. J. S. Dias,  (2004). {\em Phys. Rev. D} {\bf 70}, 084011.
\bibitem{twenty three}
V. Cardoso, O. J. S. Dias and S. Yoshida, (2006). {\em Phys. Rev. D} {\bf 74}, 044008.
\bibitem{twenty four}
S. W. Hawking, C. J. Hunter and M. M. Taylor-Robinson, (1999). {\em Phys. Rev. D} {\bf 59}, 064005.
\bibitem{twenty five}
B. Carter, (1973). {\em Black Holes (les Astres Occlus)}
(C. Dewitt and B.C. Dewitt, eds.)
New York: Gordon and Breach.
\bibitem{twenty six}
Z. Stuchlik, (1983). {\em Astron. Inst. Czechosl} {\bf 34}, 129.
\bibitem{twenty seven}
Damour, T. (1997). In: {\em Proceedings of the First Marcel Grossmann Meeting on
General Relativity}, North-Holland, Amsterdam, p. 476.
\bibitem{twenty eight}
A. N. Aliev and V. P. Frolov, (2004). {\em Phys. Rev. D} {\bf 69}, 084022.
\bibitem{twenty nine}
A. N. Aliev and C. Saclioglu, (2006). {\em Phys. Lett. B} {\bf 632}, 725.
\bibitem{thirty}
A. N. Aliev, (2007). {\em Electromagnetic Properties of Kerr-anti-de Sitter Black Holes}, arXiv:hep-th/0702129v3.
\bibitem{thirty one}
Zhao Zheng and Dai Xianxin, (1991). {\em Chin. Phys. Lett.} {\bf 8}, 548.
\bibitem{thirty two}
 C. A. Li,  (2000). {\rm Acta Phys. Sin.} {\bf 49} 1648 (in Chinese).
\bibitem{thirty three}
X. Zhang, G. Z. Xie, G. Zhao, L. Ma  and J. M. Bai,  (2000). {\rm Acta Phys. Sin.} {\bf 49} 379 (in Chinese).
\bibitem{thirty four}
J. L. Jing, (2000). {\rm Chin. Phys.} {bf 9} 389.
\bibitem{thirty five}
J. L. Jing, (2000). {\rm Chin. Phys.} {bf 10} 233.
\bibitem{thirty six}
S. C. Xie, X. T. Yang, S. Z. Yang and L. B. Lin, (2001). {\rm Chin. Phys.} {bf 10} 979.
\bibitem{thirty seven}
F. Z. Peng and  L. Liu,  (1999). {\rm Acta Phys. Sin.} {\bf 48} 6 (in Chinese).
\bibitem{thirty eight}
Z. H. Li and  L. Q. Mi,  (1999). {\rm Acta Phys. Sin.} {\bf 48} 575 (in Chinese).
\bibitem{thirty nine}
J. L. Lu and  Y. J. Wang,  (1999). {\rm Acta Phys. Sin.} {\bf 48} 389 (in Chinese).
\bibitem{fourty}
G. Chen, (1999). {\rm Acta Phys. Sin.} {\bf 48} 992 (in Chinese).
\bibitem{fourty one}
Z. Zhao, S. Y. Pei and  L. Liu,  (1999). {\rm Acta Phys. Sin.} {\bf 48} 2004 (in Chinese).
\bibitem{fourty two}
W. B. Liu and  X. Li,  (1999). {\rm Acta Phys. Sin.} {\bf 48} 1793 (in Chinese).
\bibitem{fourty three}
J. Y. Zhang,  (1999). {\rm Acta Phys. Sin.} {\bf 48} 2158 (in Chinese).

\end{thebibliography}
\end{document}